%% file: main.tex
\documentclass{IEEEtran} 
\usepackage[utf8]{inputenc}
\usepackage[T1]{fontenc}
\usepackage{microtype}
\usepackage{xspace}
\usepackage{cite}
\usepackage{array}
\usepackage{fancyvrb}
\usepackage{color}
\usepackage{colortbl}
\usepackage{caption}
\usepackage{subfig}
\usepackage{dcolumn}
\usepackage{paralist}
\usepackage{multirow}
\usepackage{listings}
\usepackage{hyperref}
\usepackage{url}
\usepackage{graphicx}

\usepackage{dcolumn}
\usepackage{flushend}

\newcommand{\smallsec}[1]{\smallskip\noindent{\bf #1.}}

\renewcommand{\paragraph}{\smallsec}

\newcommand\code[1]{{\footnotesize\textsf{#1}}}
\newcommand\cc{\textsc{CPAchecker}\xspace}

\newcommand{\Bool}		{$Bool${}\xspace}
\newcommand{\IntEq}		{$IntEq\setminus Bool${}\xspace}
\newcommand{\IntEqAdd}	{$IntEqAdd\setminus IntEq${}\xspace}
\newcommand{\Int}			{$Int\setminus IntEqAdd$\xspace}

\begin{document}

\lstdefinelanguage{ar}[ANSI]{c++}%
  {morekeywords={introduction,shadow,automaton,fail,refines,super,Super,original,this,layer,pointcut,call,aspect,execution,advice,around,before,after,execution,this,target,within,args,declare,parents,implements,throws,returning,throwing,throw,synchronized,boolean,proceed,cflow,cflowbelow,null,abstract,extends,interface,instanceof,privileged,cclass,document,grammar,overrides},%
   sensitive=f
   }[keywords,comments,strings]%
  
\lstset{basicstyle=\sf\footnotesize,
	keywordstyle=\sf\footnotesize\bfseries,
	commentstyle=\sf\footnotesize\it,
	escapechar=,
	tabsize=2,
	language=ar,
	numbers=left,
	firstline=1,
	rulesepcolor=\color{black},
	backgroundcolor=\color{white},
	numbersep=5pt,
	xleftmargin=7.5pt,
	xrightmargin=7.5pt,
	firstline=1,
	frame=top|bottom|left|right,
	showstringspaces=false,
	mathescape=true,
	fancyvrb=true,
	columns=fullflexible,
	escapeinside={(*@}{@*)},
	moredelim =**[is][\color{red}]{RED@}{@},
	moredelim =**[is][\color{blue}]{BLUE@}{@},
	moredelim =**[is][\color{darkgreen}]{GREEN@}{@},
	moredelim =**[is][\color{purple}]{PURPLE@}{@},
}

\long\def\alex#1{% draft comment
   \strut\nobreak%
  \textbf{Alex}\nobreak{\bfseries [#1]}
  }

\newcommand{\cpachecker}{{\small\scshape CPAchecker}\xspace}
\newcommand{\cpacheckerabe}{{\small\scshape CPAchecker 1.0.10-abe}\xspace}
\newcommand{\cpacheckermemo}{{\small\scshape CPAchecker 1.0.10-memo}\xspace}
\newcommand{\cpacheckerexplicit}{{\small\scshape CPAchecker 1.1.10-Explicit}\xspace}
\newcommand{\cpacheckerseqcom}{{\small\scshape CPAchecker 1.1.10-SeqCom}\xspace}
\newcommand{\cpacheckerexplicits}{{\small\scshape CPAchecker-Explicit}\xspace}
\newcommand{\cpacheckerseqcoms}{{\small\scshape CPAchecker-SeqCom}\xspace}
\newcommand{\cseq}{{\small\scshape CSeq 2012-10-22}\xspace}
\newcommand{\blast}{{\small\scshape Blast}\xspace}
\newcommand{\blastpl}{{\small\scshape Blast}\xspace}
\newcommand{\slam}{{\small\scshape Slam}\xspace}
\newcommand{\magic}{{\small\scshape Magic}\xspace}
\newcommand{\cbmc}{{\small\scshape Cbmc}\xspace}
\newcommand{\fshell}{{\small\scshape FShell 1.3}\xspace}
\newcommand{\satabs}{{\small\scshape SATabs 3.0}\xspace}
\newcommand{\sycmc}{{\small\scshape SyCMC}\xspace}
\newcommand{\predator}{{\small\scshape Predator}\xspace}
\newcommand{\wolverine}{{\small\scshape Wolverine 0.5c}\xspace}
\newcommand{\llbmc}{{\small\scshape Llbmc 2012-10-23}\xspace}
\newcommand{\esbmc}{{\small\scshape Esbmc 1.20}\xspace}
\newcommand{\qarmc}{{\small\scshape QArmc-Hsf(c)}\xspace}
\newcommand{\threader}{{\small\scshape Threader 0.92}\xspace}
\newcommand{\ufo}{{\small\scshape Ufo 2012-10-22}\xspace}
\newcommand{\ultimate}{{\small\scshape Ultimate 2012-10-25}\xspace}
\newcommand{\symbiotic}{{\small\scshape Symbiotic 2012-10-21}\xspace}

\newcommand{\csisat}{{\small\scshape CSIsat}\xspace}
\newcommand{\cvc}{{\small\scshape Cvc}\xspace}
\newcommand{\mathsat}{{\small\scshape MathSAT}\xspace}
\newcommand{\simplify}{{\small\scshape Simplify}\xspace}
\newcommand{\picosat}{{\small\scshape PicoSAT}\xspace}
\newcommand{\clpprover}{{\small\scshape CLPprover}\xspace}
\newcommand{\foci}{{\small\scshape Foci}\xspace}
\newcommand{\sicstus}{{\small\scshape SICStus Prolog}\xspace}
\newcommand{\moped}{{\small\scshape Moped}\xspace}
\newcommand{\buddy}{{\small\scshape BuDDy}\xspace}
\newcommand{\javabdd}{{\small\scshape JavaBDD}\xspace}

\newcommand{\cil}{{\small\scshape Cil}\xspace}
\newcommand{\eclipse}{{\small\scshape Eclipse}\xspace}

\newcommand{\mytitle}{Domain Types:\\ Selecting Abstractions Based on Variable Usage}
\newcommand{\myauthors}{
Sven Apel\,$^1$,
Dirk Beyer\,$^1$,
Karlheinz Friedberger\,$^1$,
Franco Raimondi\,$^2$,
and Alexander von Rhein\,$^{1}$
}
\newcommand{\myaffiliations}{
{$^1$\,University of Passau, Germany}\\
{$^2$\,Middlesex University, London, UK}
}

\pagestyle{empty}
\hspace{1.5mm}
\begin{minipage}{17cm}
\begin{center}
~\\[3cm]
\Huge{\mytitle}
\\[2cm]
\large{\myauthors}
\\[1cm]
\normalsize
	\myaffiliations\\[5cm]

\vspace{20mm}
\hspace{-5mm}
\includegraphics[scale=0.2]{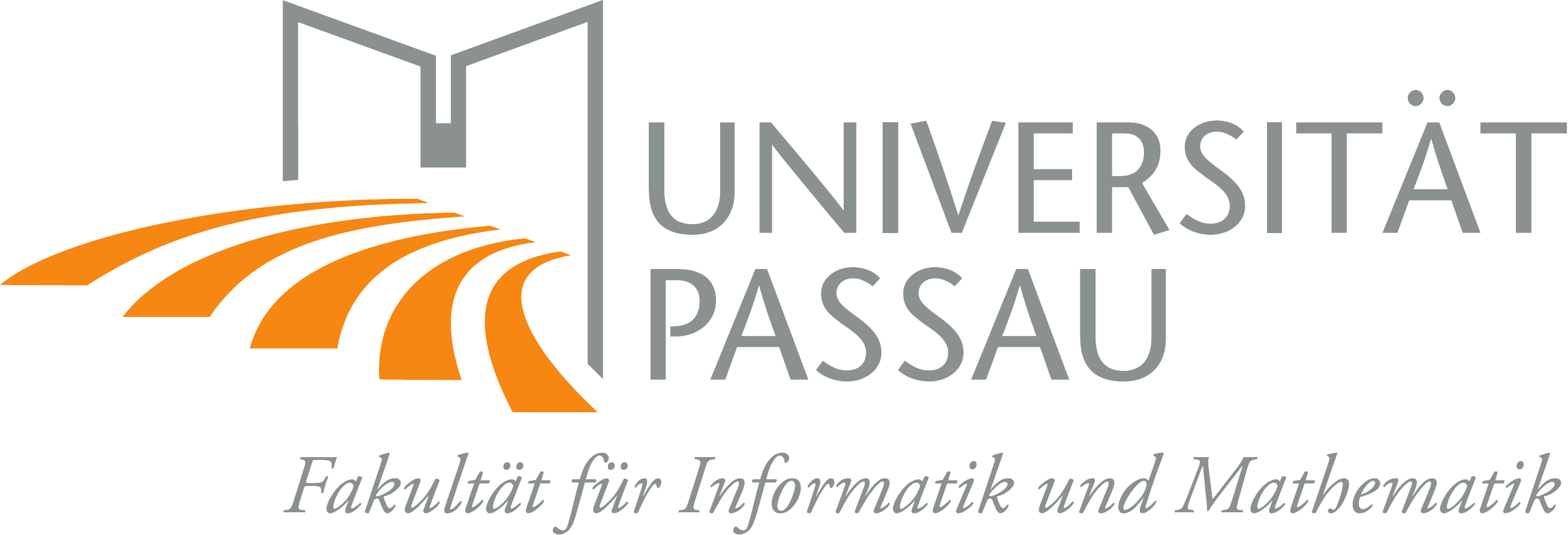} \\[1cm]
Technical Report, Number MIP-1303\\
Department of Computer Science and Mathematics\\
University of Passau, Germany\\
May 2013
\end{center}
\end{minipage}

\title{\mytitle}

\author{
\myauthors
\vspace{1ex}\\
\myaffiliations
}

\begin{titlepage}
\null
\end{titlepage}

\maketitle

\pagenumbering{arabic}
\thispagestyle{empty}
\pagestyle{plain}

\begin{abstract}
\input{sections/abstract}
\end{abstract}

\input{sections/introduction}

\input{sections/background}
\input{sections/approach}

\input{sections/evaluation}

\input{sections/related}

\input{sections/conclusion}
\bibliographystyle{plain}
\bibliography{domaintypes,sw,dbeyer}
\end{document}

%% file: sections/abstract.tex
The success of software model checking depends on
finding an appropriate abstraction of the subject program.
The choice of the abstract domain and the analysis configuration is currently left to the user, 
who may not be familiar with the tradeoffs and performance details of the available abstract domains.
We introduce the concept of \emph{domain types}, which
classify the program variables into types that are more fine-grained
than standard declared types, such as \code{int} or \code{long}, in order
to guide the selection of an appropriate abstract domain for a model checker.
Our implementation determines the domain type for each variable in a pre-processing step,
based on the variable usage in the program,
and then assigns each variable to an abstract domain.
The model-checking framework that we use supports to specify a separate analysis precision 
for each abstract domain, such that we can freely configure the analysis.
We experimentally demonstrate a significant impact of the choice of the abstract domain
per variable.
We consider one explicit (hash tables for integer values) and
one symbolic (binary decision diagrams) domain.
The experiments are based on standard verification tasks that are taken from recent
competitions on software verification.
Each abstract domain has unique advantages in representing the state space
of variables of a certain domain type.
Our experiments show that software model checkers can be improved 
with a domain-type guided combination of abstract domains.

%% file: sections/introduction.tex
\section{Introduction}

One of the main challenges in software model checking is to
automatically find, for each program variable, the right abstract
representation (also known as \emph{abstract domain}) that suffices to efficiently
prove the program correct or to identify an error path.  
Several abstract domains have been applied successfully to software-verification
problems, with different strengths and weaknesses.\,%
\footnote{For an overview, we refer the reader to the competition on software verification: \href{http://sv-comp.sosy-lab.org/}{{\tt http://sv-comp.sosy-lab.org}}\,.}
Abstract domains can be based on  
explicit (e.g., hash tables for integers, memory graphs for the heap) and 
symbolic (predicates, binary decision diagrams (BDD)) representations.
For example, 
using an explicit-value domain~\cite{CPAexplicit} was efficient on many
benchmarks from the recent competition on software verification, 
while using a BDD domain~\cite{CPABDD} was more efficient on event-condition-action (ECA)
systems that involve only simple operations over integers in an ECA competition~\cite{RERS12}.
In the context of product-line verification, it has been shown that
BDD-encodings of feature variables
improve the performance~\cite{FAV_ICSE13,ClassenBDD}.  
The overall picture is that different abstract domains are successful on different
types of programs, and for every abstract domain, we can find programs
for which the abstract domain is not successful.

So far, the choice of the abstract domain for a given
verification problem (which often implies the choice of a certain
verification tool as well) was left to the user.
The precision (determining which facts to track) for an analysis using a particular abstract domain
is often automatically adjusted using counterexample-guided abstraction refinement~\cite{ClarkeCEGAR}.
Also, several (component) analyses can be combined to an analysis combination,
where each component analysis has its own precision~\cite{CPAplus,CPAexplicit}, 
which can be dynamically adjusted (based on certain domain-dependent measures) and
determines the level of abstraction inside the component analysis.
This concept is used to switch to another domain if the current domain is not successful.

Our goal is to automate the choice of an effective abstract domain using
a pre-analysis (that runs before the model checker starts the state-space exploration)
and to automatically assign program variables to abstract domains.  
To achieve this goal, we analyze the
usage of program variables and assign each variable to a certain
domain type.  In addition to the standard declared type in the programming language
(e.g., \code{int}, \code{char}, \code{bool}), the \emph{domain type} of a variable
represents information about the value range and the operations in which the variable is involved.
Using this idea, we can even determine domain types for
variables in dynamically typed, or untyped, languages
(we focus on statically-typed C programs, though).

Our approach is based on a verification framework in which each
abstract domain has a \emph{precision} associated with it~\cite{CPAplus}. 
We can now use the domain types from the pre-analysis as guidance for
assigning an abstract domain to each variable.
In the experiments that we conducted to demonstrate the impact of our idea,
we use two abstract domains, namely an explicit-value domain
and a BDD-based domain.
For both domains, the precision is a set of variables (which shall be
tracked in the domain).  
Depending on the domain type, 
we add each variable to the precision of either the explicit-value domain or
the BDD domain.
The precision of the abstract domain instructs the analysis to track only those
variables with that abstract domain that occur in its precision.
If the domain assignment is good, then this approach improves the
overall verification performance, because then each domain manages only the variables that it
is best suited for.

Our analysis is implemented in the verification framework \cc~\cite{CPACHECKER}, 
which implements configurable program analysis 
for \textsf{C} programs and provides abstract domains for
an explicit-value analysis and a BDD-based analysis.
However, the approach is generally applicable to other model checking tools, 
other abstract domains, and to other (statically or dynamically typed) 
programming languages that support different domain types.

We evaluate our approach on 7 benchmark sets from different application domains 
(a total of 335 files) that have been used by recent international competitions 
on software model checking (SV-COMP 2012, 
RERS challenge 2012~\cite{SVCOMP12,RERS12}).
We explore different mappings from domain types to abstract domains 
and discuss which abstract domain proves suitable for which domain type.
We also compare our approach to a competitive model-checking tool that 
won several awards in software model-checking competitions.

Our evaluation shows that the programs in our benchmark sets contain a significant number 
of variables that have a much narrower domain type than the declared type of the variable.
The evaluation also shows that the performance of model checking improves when these variables 
are analyzed with a suitable abstract domain.
Our results are available on the supplementary 
website\,\footnote{\url{http://www.sosy-lab.org/projects/domaintypes/}}.

\begin{figure}[t]
\begin{lstlisting}[numbers=none,frame=top|bottom,basicstyle=\sf\scriptsize,
keywordstyle=\sf\scriptsize\bfseries,commentstyle=\sf\scriptsize\it]{ar}
int enabled, a, b;
b=20;
if (enabled || a > 5) {
  if (a == 0) {
    b = 0;
  }
  assert (b*b > 200);
}
\end{lstlisting}
%\vspace{-3mm}
\caption{Example with variables of different domain types}
\label{fig:domainVariables}
%\vspace{-4mm}
\end{figure}

\smallsec{Example} Fig.~\ref{fig:domainVariables} illustrates the
advantage of our approach on an example program.  The program contains
three variables that are declared by the programmer as `integer'.  The
variables are used in different ways: the variable \code{enabled} is
used as a boolean and the variables \code{a} and \code{b} are numeric;
variables \code{a} and \code{b} are used in a greater-than comparison
and \code{b} is also used in a multiplication.  Neither the
explicit-value analysis nor the BDD-based analysis is able to
efficiently verify such programs: The explicit-value domain is
perfectly suited to handle variable \code{b}, because \code{b} has a
concrete value, and the multiplication and the greater than comparison
can be easily computed, whereas BDDs are known to be inefficient for
multiplications~\cite{McMillan97}.  The BDD domain can efficiently
encode the variables \code{enabled} and \code{a}, whereas the
explicit-value analysis is not good at encoding facts like~$a > 5$.
Thus, without the information about variable~\code{a}, the
explicit-value analysis does not know the value of variable~\code{b}
and cannot determine the result of the multiplication.

A solution that has been proposed before is to use both abstract 
domains in parallel, 
with each domain handling all variables.
If the domains are well communicating (reduced products), this could solve the verification task,
but the load on each domain would be unnecessarily high, because 
every domain has to handle more variables.
Our experiments show that the load on the abstract domain should not be underestimated, 
especially considering the BDD domain.

\pagebreak
\smallsec{Contributions} We make the following novel contributions:
\begin{compactitem}
\item We developed the concept of domain types and designed
       a pre-analysis that determines domain types of program variables.
\item We extended an existing model-checking tool to use and synchronize several abstract 
       domains (explicit-value and BDD) in parallel, while each domain handles 
       only the variables that it is suited for.
\item We evaluate our approach on seven benchmark sets from 
       competitions on software verification.
\end{compactitem}

%% file: sections/background.tex
\section{Background}

We explain the concepts that we use in our work informally,
and give references below to the literature for detailed descriptions.
In the presentation, whenever a concrete context is necessary,
we assume to verify C programs, and that we analyze integer variables.

\paragraph{Abstract Domains and Program Analysis}
Abstraction-based software model checkers automatically 
extract an abstract model of the subject program
and explore this model using one or more abstract domains.
An abstract domain is an abstract representation of certain aspects of the concrete program 
states that the state exploration is supposed to track~\cite{DragonBook}.
Different abstract domains can track different aspects of the program state space and complement
each other.
For example, a \emph{shape domain} stores, for each tracked pointer,
the shape of the pointed-to data structures on the heap~\cite{TVLA,ShapeRefine,PREDATOR-COMP}.
Another example is the \emph{explicit-value domain} that, for each tracked variable,
tracks the explicit value of the variable~\cite{Pathfinder,CPAexplicit,SPIN}.

The two examples illustrate that abstract domains can represent different information.
However, it is also possible to use different abstract domains
to represent the same information in different ways.
For example, consider a program in which the value of variable~$x$ ranges from~$3$ to~$9$.
This can be stored by an \emph{interval domain} using the abstract state~$x \mapsto [3, 9]$~\cite{ASTREE},
or by a \emph{predicate domain} using the abstract state~$x \geq 3 \land x \leq 9$~\cite{GrafSaidi97,SLAM,BLAST}.

Every abstract domain consists of 
\begin{compactitem}
\item[(a)] a representation of sets of concrete states, defining the abstract states (lattice elements),
\item[(b)] an operator to decide if one abstract state subsumes another abstract state (partial order), and
\item[(c)] an operator that combines two abstract states into a new abstract state that represents both (join).
\end{compactitem}
Every tool for program analysis uses one or several abstract domains to represent
the states of the program.
The abilities of the abstract domain imply the effectivity (is the analysis able to correctly solve the
verification problem?) and efficiency (is the analysis fast, does it scale to large programs?)
of the program analysis.

\paragraph{Precision}
Each abstract domain can operate on different levels of abstraction,
i.e., it can be more fine-grained or more coarse.
The level of abstraction of an abstract domain is determined by the \emph{abstraction precision},
which controls if the analysis is coarse or detailed.
For example, the precision of the shape domain could instruct the analysis which
pointers to track, and how large a shape can maximally grow.
The precision of the predicate domain is a set of predicates to track,
which can, for example, grow by adding predicate during refinement steps~\cite{ClarkeCEGAR}.

Next, we describe the two abstract domains that we consider in our experiments.

\begin{figure}[t]
\vspace{-8mm}
\subfloat[C Code]{
\begin{minipage}[b][1\width]{0.2\textwidth}
\centering
\begin{lstlisting}[firstnumber=0,frame=top|bottom,basicstyle=\sf\scriptsize,
keywordstyle=\sf\scriptsize\bfseries,commentstyle=\sf\scriptsize\it,language=ar,linewidth=\textwidth]
int x,y,z;
x=5;
if (y > 1) {
	z = 2;
} else {
	z = 3-1;
}
\end{lstlisting}
\end{minipage}
}
\hfill
\subfloat[Abstract reachability graph]{%trim=l b r t
\begin{minipage}[b][1\width]{0.3\textwidth}
\centering
\includegraphics[trim=0cm 7cm 6cm 0cm, clip=true, width=30mm]{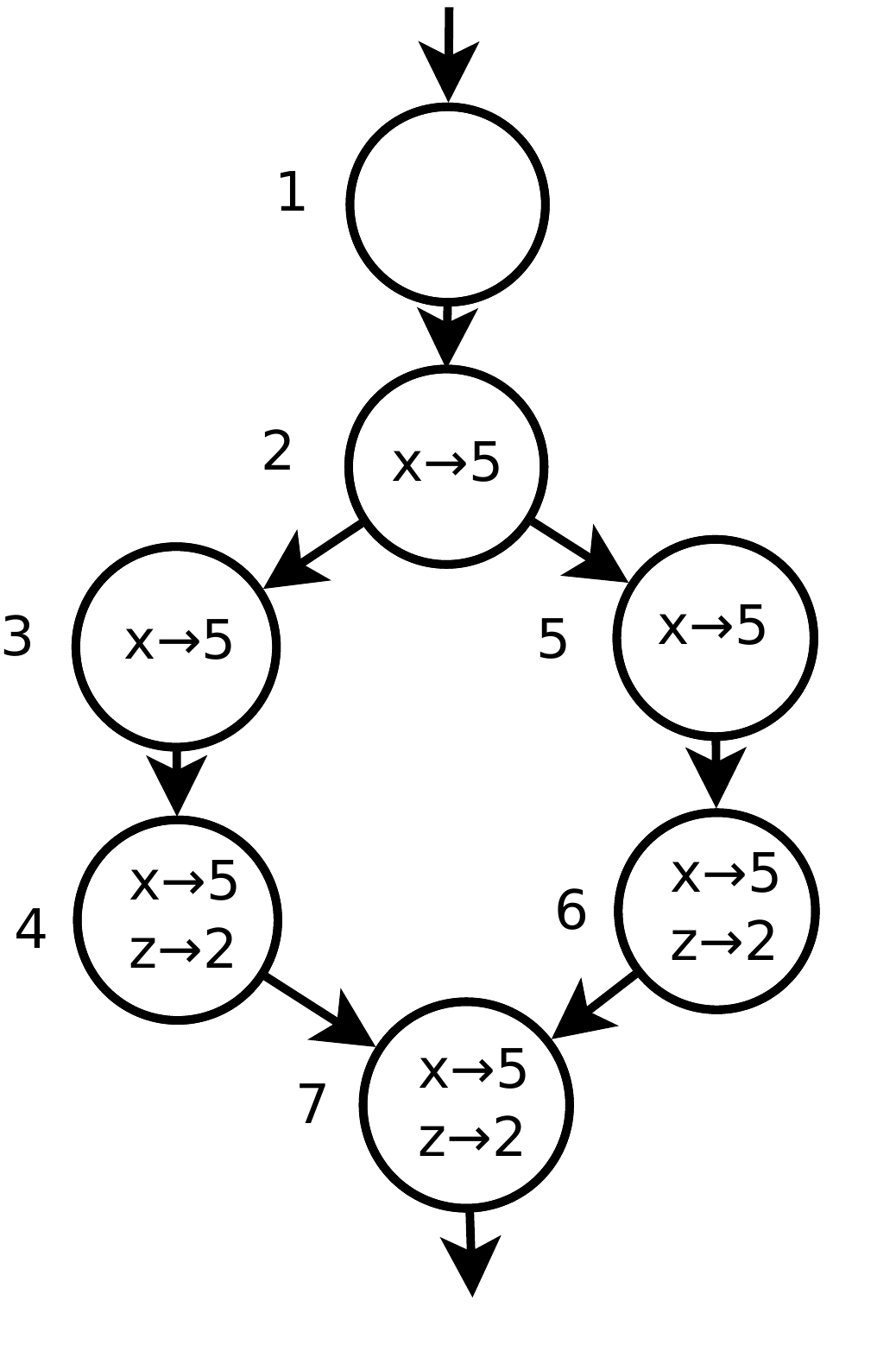}
\end{minipage}
}
\caption{Code example and abstract reachability graph (nodes represent explicit states)}
\label{fig:exampleExplicit}
\end{figure}

\paragraph{Explicit-Value Domain}
The explicit-value domain stores concrete values for all program variables 
that occur in the precision (once a concrete value has been determined).
Each abstract state of the domain is represented by a map that assigns to each program variable
an integer value.
Variables for which a concrete value cannot be determined do not appear in the map.
For example, consider the code in Fig.\,\ref{fig:exampleExplicit}~(a) 
and the corresponding abstract reachability graph (ARG) in 
Fig.\,\ref{fig:exampleExplicit}~(b):
Variable \code{x} is assigned, and the value is stored in a new abstract state 
(state 2 in Fig.\,\ref{fig:exampleExplicit}~(b)).
Then, a conditional statement spawns two possible execution paths, so the model 
checker explores both paths.
The explicit domain cannot store information on variable \code{y}, 
because it does not have a concrete value.
After both branches of the conditional statement are executed, the ARG has two 
``frontier'' abstract states that are identical.
Only one of these abstract states is stored because the other is subsumed,
and state exploration continues from this `merged' state.

The explicit-value domain might suffer from a loss of information 
in cases where not all information can be stored, e.g., \code{ y > 1}.
On the one hand, this introduces imprecision and potentially false alarms, 
but, on the other hand, if values are present, all operations can be 
executed extremely fast.

The abstraction precision controls which variables should be tracked in the explicit-value domain. 
For the code fragment from Fig.~\ref{fig:exampleExplicit},
we could use a precision~$\{\code{x}, \code{z}\}$ and omit~\code{y},
if we knew beforehand that there is no point in representing variable~\code{y}
in the explicit-value domain.

\paragraph{BDD Domain}
The BDD domain stores information about program variables using binary decision diagrams (BDD).
Each abstract state in the BDD domain is represented by a predicate over the variable values
that the BDD represents~\cite{Bryant1992}.
BDDs can be efficient in storing predicates and performing boolean operations.
Because of this characteristics, BDDs have been used in model checking of 
systems with a large number of boolean variables, most prominently in 
hardware verification~\cite{McMillan97,SymbolicModelChecking}.

Values of integer variables can be represented by BDDs
using a binary encoding of the values and representing the binary values in 
32~boolean BDD variables.
In our example, this would require 96~BDD variables for the three integer program variables.
Because the size of the BDD has a strong impact on the performance of BDD operations, 
it is important to keep the number of BDD variables small.

The abstraction precision of the BDD domain is also a simple set of program variables
that the analysis should track using this abstract domain.
Considering again our example, if we knew beforehand that variables~\code{x} and~\code{y}
can be efficiently represented by the explicit-value domain,
we would not include them in the precision, which would result in precision~$\{\code{z}\}$
for the BDD domain, which needs only 32~BDD variables.

The additional power that allows us to store disjunctions in BDDs
comes at a price: the performance decreases with a growing number of variables,
and thus, we should parsimoniously use the BDD domain.

To achieve the goal of a better assignment of variables to abstract domains,
we introduce the concept of domain types in the next section.

%% file: sections/approach.tex
\section{Domain Types}
\label{sec:approach}
Our new approach performs the verification process in three steps:
(1)~We start with analyzing the subject program in order to determine 
the domain type for each variable (pre-analysis).
(2)~Then, each variable is mapped to an abstract domain that the 
analysis will use to represent information about the variable.
The mapping from domain type to abstract domain is not yet automated.
This mapping determins the precision for each abstract domain.
(3)~Finally, the actual model-checking procedure with the individual precisions
per abstract domain is started.

\begin{figure}[t]
\vspace{2mm}
\includegraphics[trim=0cm 11cm 2cm 2cm, clip=true, width=0.9\columnwidth]{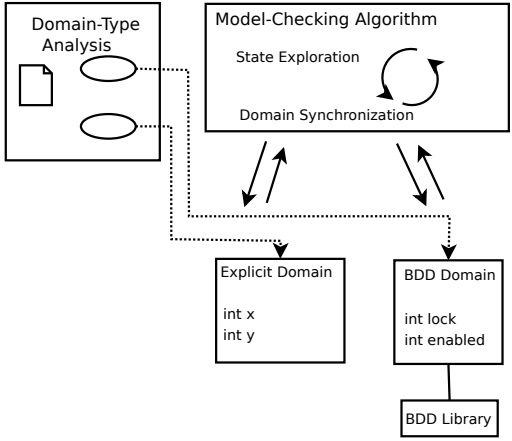}
\caption{A model-checking engine with two abstract domains and domain-type analysis.}
\label{fig:overviewModelchecking}
\end{figure}

Figure\,\ref{fig:overviewModelchecking} illustrates our approach of a 
model-checking engine that is based on domain-types.
The state exploration algorithm is implemented in the main module.
It uses several abstract domains to represent the state space of the program.
Note that each variable is tracked by only one abstract domain.

\subsection{Domain Types --- Classification}
In statically-typed programming languages, each variable is declared to be of a certain type.
The type determines which values can be stored in the variable and which operations
can use the variable as operand.
For the assignment of abstract domains to variables,
we need more specific information about the variables,
in particular, information on the operations that the variables are involved in.

For example, consider boolean variables in the programming language~\textsc{C}.
The language \textsc{C} does not provide a type `boolean'.
In \code{C}, the boolean values \code{true} and \code{false}
are represented by the integer values \code{1} and \code{0}, respectively.
When integer variables are read, the value \code{0} is interpreted 
as \code{false} and all other values are interpreted as \code{true}.
Let us consider the tiny code fragment in Fig.\,\ref{fig:booleanC}).

\begin{figure}[t]
\vspace{6mm}
\begin{lstlisting}[numbers=none,firstnumber=1,frame=top|bottom,basicstyle=\sf\scriptsize,
keywordstyle=\sf\scriptsize\bfseries,commentstyle=\sf\scriptsize\it]{ar}
int enabled;
...
if (enabled) {
	...
} else {
	...
}
\end{lstlisting}
\caption{Using an integer variable as boolean in \textsc{C}}
\label{fig:booleanC}
\end{figure}

The expression \code{enabled} in the \code{if} condition is internally expanded
to the expression \code{enables != 0}~\cite{ANSI:1999:AII}.
It is clear from the last section that such a variable should be represented in a BDD
by one single boolean variable, and not 32 boolean variables.
Therefore, we introduce a domain type \emph{Bool} that represents the more precise type.
To determine whether an integer variable has actually the domain type \emph{Bool}, 
our pre-analysis inspects all occurrences of the variable in expressions.
If a variable is found to be of domain type \emph{Bool}, this fact can be considered
in the assignment of the abstract domain and thus, the variable can be represented
by a more efficient data structures during model checking.

Other programming languages such as \textsc{Java} provide more restrictive types 
like \code{boolean} and \code{byte},
but for the purpose of assigning the best abstract domain,
more precise information is beneficial.
In dynamically-typed or even untyped languages, types of variables are unknown 
before program execution.
A static analysis of domain types can lead to considerable improvements here, 
because it can infer quite constrained domain types.
This information can be used during the verification to choose efficient 
algorithms and data structures.

Figure~\ref{fig:types} shows the four domain types that we consider in the static pre-analysis 
(many more are possible to be explored in future work).
Our pre-analysis assigns every program variable to exactly one of these domain types,
from which an appropriate abstract domain can be derived.

\begin{table}[t]
\caption{Domain types that are considered in this paper}
\label{fig:types}
\vspace{2mm}
\centering
\begin{tabular}{ll}
  \hline
  Domain type               & Short description\\
  \hline
  {\sf \textit{Bool}}      & Boolean variable\\
  {\sf \textit{IntEq}}     & Equality comparisons with \\
                             &  a constant value (\code{==},\code{!=})\\
  {\sf \textit{IntEqAdd}}  & Linear arithmetics only (\code{+,--})\\
  {\sf \textit{Int}}       & All integer variables\\
  \hline
\end{tabular}
\end{table}

\pagebreak
\subsection{Domain Types --- Analysis}
The first step of our approach is a static domain-type analysis that determines 
the domain types for all program variables.
It works on an abstract representation of the program (a control-flow graph, 
in our case, but an abstract syntax tree would also be sufficient) and processes 
all program operations in which variables are used.
For a variable~\code{x}, the analysis in its search analyzes all expressions in which
the variable \code{x} is involved.
For example, the expressions \code{x==0}, \code{x==x+1}, and \code{y==x*z}
yield the domain types \emph{Bool}, \emph{IntEqAdd}, and \emph{Int}, respectively.

\begin{figure}[t]
\centering
\includegraphics[trim=0cm 14cm 0cm 0cm, clip=true, width=0.7\columnwidth]{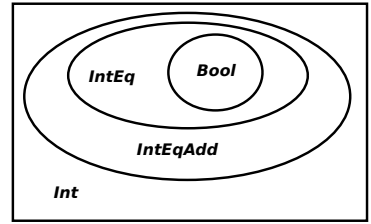}
\caption{Hierarchical structure of domain types}
\label{fig:usageTypesStructure}
\vspace{5mm}
\end{figure}

In the following, we informally define the criteria that the type analysis
is based on (for brevity, we omit the formal type system).
The four domain types are hierarchically overlapping (subtypes),
as illustrated in Fig.~\ref{fig:usageTypesStructure}.

\paragraph{Bool}
A variable belongs into the domain type \emph{Bool} if all expressions in 
which the variable is used are the
boolean expressions \code{\&\&}, \code{||}, and \code{!}, as well as
the comparisons with zero \code{==0} and \code{!=0}.
Comparisons with other variables are possible if their domain type is \emph{Bool} as well.

\paragraph{IntEq}
Variables of domain type \emph{IntEq} are limited to boolean expressions, equality tests with 
constant values and simple comparisons (\code{==},\code{!=}) 
with other variables of domain type \emph{IntEq}.

\paragraph{IntEqAdd}
The domain type \emph{IntEqAdd} contains all variables that are used in 
boolean expressions (\code{\&\&}, \code{||}, \code{!}),
linear arithmetic (\code{+},\code{--}), 
comparisons (\code{==},\code{!=},\code{<},\code{>},\code{<=},\code{>=}), 
and in bit operations (\code{\&},\code{|},\code{\^}).

\paragraph{Int}
All other variables are of domain type \emph{Int}.
This includes variables that use multiplication (\code{*}), 
division (\code{/}), and bitshift operations (\code{$<<$}, \code{$>>$}).

Each variable that is of type \emph{Bool}, is also of types \emph{IntEq}, \emph{IntEqAdd}, 
and \emph{Int}.
The type system assigns the strongest (most restrictive) possible type,
i.e.\ the type system prefers domain type \emph{Bool} over \emph{IntEqAdd}).
When we discuss variables of domain types in the text, we usually want to 
refer only to variables that are not of the ``smaller'' domain types 
(e.g.,~only the variables in \emph{IntEq} that are not \emph{Bool});
we use the set notation to denote this (e.g., \IntEq).
Fig.\,\ref{fig:exampleIntEqualUsageType} shows three variables of domain type \emph{IntEq}.
Variable \code{a} for example is (more precisely) of domain type \emph{Bool}.
In such cases, we assign the most restricted domain type: \emph{Bool}.
\begin{figure}[t]
\begin{lstlisting}[numbers=none,firstnumber=1,frame=top|bottom,basicstyle=\sf\scriptsize,
keywordstyle=\sf\scriptsize\bfseries,commentstyle=\sf\scriptsize\it,language=ar]
if (a == 0) {
	b = 1042;
	c = b;
} else {
	c = 989;
}
\end{lstlisting}
\caption{Example of variables of \emph{IntEq} domain type}
\label{fig:exampleIntEqualUsageType}
\end{figure}

We consider usual optimizations, for example,
if \emph{IntEq} variables are limited to a small set of values,
we use this information to re-map the possible values to a simpler domain of successive numbers
and possibly save boolean BDD variables if the domain type \emph{IntEq} gets assigned to 
the BDD domain.

\subsection{Domain Assignment}

Once the domain types have been determined for all variables, we need to 
assign each variable to a certain abstract domain that the analysis
uses to track the variable.
For this, we use the \emph{domain assignment}, which is 
a map that assigns an abstract domain to each domain type.
For each abstract domain, we add all variables that the abstract domain should track
to the abstraction precision of that abstract domain.
In principle, every abstract domain can represent any variable,
but each abstract domain has certain strengths and weaknesses
(in terms of effectiveness and efficiency).
Therefore, we have to map every domain type to an abstract domain that is 
\emph{appropriate} for the usage of the variables.

It seems straightforward to associate the domain type \Bool with the BDD domain.
The BDD domain can efficiently represent complicated boolean combinations
of variables, but is sensitive to the number of represented variables.

We can also represent the domain types \IntEq and \IntEqAdd  by the BDD domain.
For domain type \IntEq, we know from the properties of the domain type 
that those variables only hold a limited and known set of values.
Therefore, we can enumerate these values and represent them by BDD variables.
To represent values of a set of size~$n$, we need $\log_2(n)$ BDD variables.
This representation can be very efficient.

The explicit-value domain can in principle be used for all domain types,
but the more different combinations of variable assignments need to be distinguished
in the analysis, the larger the state space grows, perhaps leading to the problem
of state-space explosion.
Also, the explicit-value domain is not well-suited for analyzing uninitialized variables.

In our experiments, we show that different domain assignments have significantly different
performance characteristics for different sets of verification tasks.
Automatically finding an optimal domain assignment remains a research problem for future work.
The goal of this paper is to show that the concepts of domain types
provides a technique to approach this problem.

%% file: sections/evaluation.tex
\pagebreak
\section{Experimental Evaluation}
\label{sec:evaluation}

To evaluate domain-type-based analysis approach, 
we conduct experiments with different configurations on 
a diverse set of verification tasks.
The results give evidence that the choice of the representing abstract domains
for the considered domain types has a significant impact on the effectiveness and efficiency.
We make the following statements:\\
\begin{compactitem}
  \item[\textbf{Domain types.}] The subject systems contain a large set of integer variables 
  that our domain-type analysis classifies into four specific domain types.
  \item[\textbf{Variable partitioning.}] The verification performance significantly 
  changes if variables are treated with different abstract domains, 
  compared to tracking all variables with the same abstract domain.
  \item[\textbf{Domain optimization.}] Using the BDD domain for variables of domain type \IntEq 
  and \IntEqAdd can improve the verification performance.
  \item[\textbf{Comparison with state-of-the-art.}] 
  The best combination of the explicit-value domain and the BDD domain with 
  domain-type analysis performs better than a competitive predicate-analysis approach 
  that uses other optimization techniques such as counterexample-guided abstraction refinement
  (CEGAR).
\end{compactitem}

\subsection{Implementation}
\label{sec:impl}
For our experiments, we extended the open verification framework \cc~\cite{CPACHECKER}.
\cc offers various abstract domains and supports the concept of abstraction precisions
in a modular way, such that it is easy to extend and configure.
The tool is applicable to an extensive set of verification
benchmarks, because it participated at the competition on software verification.
This enables us to evaluate our approach on a large set of realistic programs.
We reuse one of those abstract domains in our experiments, namely a version of the 
explicit domain (without CEGAR)~\cite{CPAexplicit}.
We extended \cc with the domain-type analysis described in the previous section, 
and we implemented a new, flexible abstract domain that uses BDDs to represent variable values.
For comparison of our domain-type-based approach with a stat-of-the-art verifier, 
we will use another configuration of \cc that participated at the competition in the last year.
This other configuration implements a predicate analysis with CEGAR, interpolation, 
and adjustable-block encoding~\cite{ABE}.

\paragraph{Explicit-Value Domain}
We use the default explicit-value domain that is already implemented in \cc.
The implementation can be used either with CEGAR or without.
Its basic version works similarly to the explicit-value domain described 
in the background section.
In this paper, we cannot compare with a CEGAR-based configuration
because abstraction refinement is orthogonal to domain-type analysis.
Therefore, we use the explicit-value domain without CEGAR in all experiments.
It is possible to apply CEGAR to the BDD-based analysis, but this is out of scope of this paper.

\pagebreak
\paragraph{BDD Domain}
The BDD domain uses binary decision diagrams (BDD) to represent the values of variables,
by encoding each integer variable with one or more (boolean) BDD variables.
The BDD domain uses a different encoding of variables in the BDD for each domain type.
For domain type \Bool, we use exactly one BDD variable per program variable.
For variables of domain type \IntEqAdd, we use 32 (boolean) BDD variables 
to represent on program variable.
For variables of domain type \IntEq, we know from the pre-analysis how many
different values the variable can hold.
Therefore, we can re-map the variable values to a new set of values with the same cardinality
and therefore need considerable fewer BDD variables (compared to 32 BDD variables). 
We use a simple bijective map from the original constants in the program
to a (smaller, successive) set of integer values, 
and need only $log_2(n)$ bits to encode a set of $n$ different program constants 
(of arbitrary values) for a program variable.
In our implementation, we use one additional boolean variable
to be able to encode the situation that the variable is different
from all program constants for the variable.

\subsection{Experimental Setup}
\label{sec:setup}
We performed all experiments on an Ubuntu~11.10 system 
with 16\,GB~RAM and an Intel i7-2600 processor (8\,cores, 3.4\,GHz).
Each verification run was limited to 15\,GB of memory and 900\,s of CPU time.
We used \cc revision 7487.
Each verification task from our benchmark set was verified using four different configurations:\\
\begin{compactitem}
\item[\textit{Explicit-Int}] This configuration tracks all variables
with the explicit-value domain.
\item[\textit{BDD-Bool}] This configuration uses both abstract domains,
where all variables of domain type \Bool are in the precision of the BDD domain
and all other variables are in the precision of the explicit-value analysis.
\item[\textit{BDD-IntEq}] This configuration uses both abstract domains,
where all variables of domain type \emph{IntEq} are in the precision of the BDD domain
and all other variables are in the precision of the explicit-value domain.
\item[\textit{BDD-IntEqAdd}] This configuration uses both abstract domains, 
where all variables of domain type \emph{IntEqAdd} are in the precision of the BDD domain
and all other variables are in the precision of the explicit-value domain.
\item[\textit{BDD-Int}] This configuration tracks all variables
with the BDD domain.
\end{compactitem}

\begin{figure*}[t!]
\centering
\includegraphics[width=\textwidth]{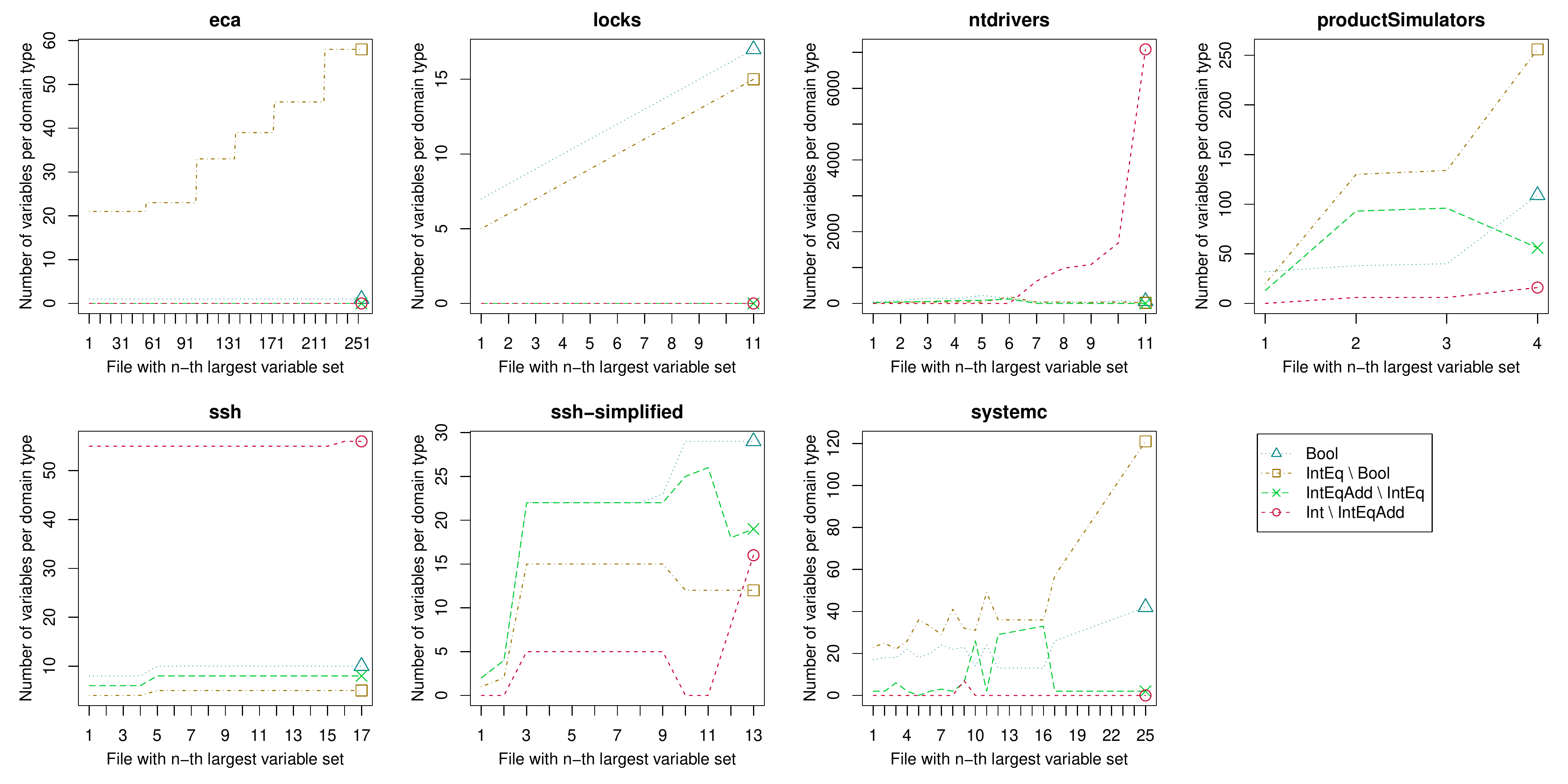}
\caption{Domain types of variables of for all verification tasks.
The diagrams show, for each benchmark set, how many variables of each domain 
type the files contain (excluding variables that are already contained in a sub-type).
The x-axis shows the verification tasks, sorted by their total number of variables and the 
y-axis shows the absolute number of variables in each domain type 
(excluding variables that are already considered for sub-types).}
\label{fig:usageTypes}
\centering
\vspace{18mm}
\includegraphics[width=\textwidth]{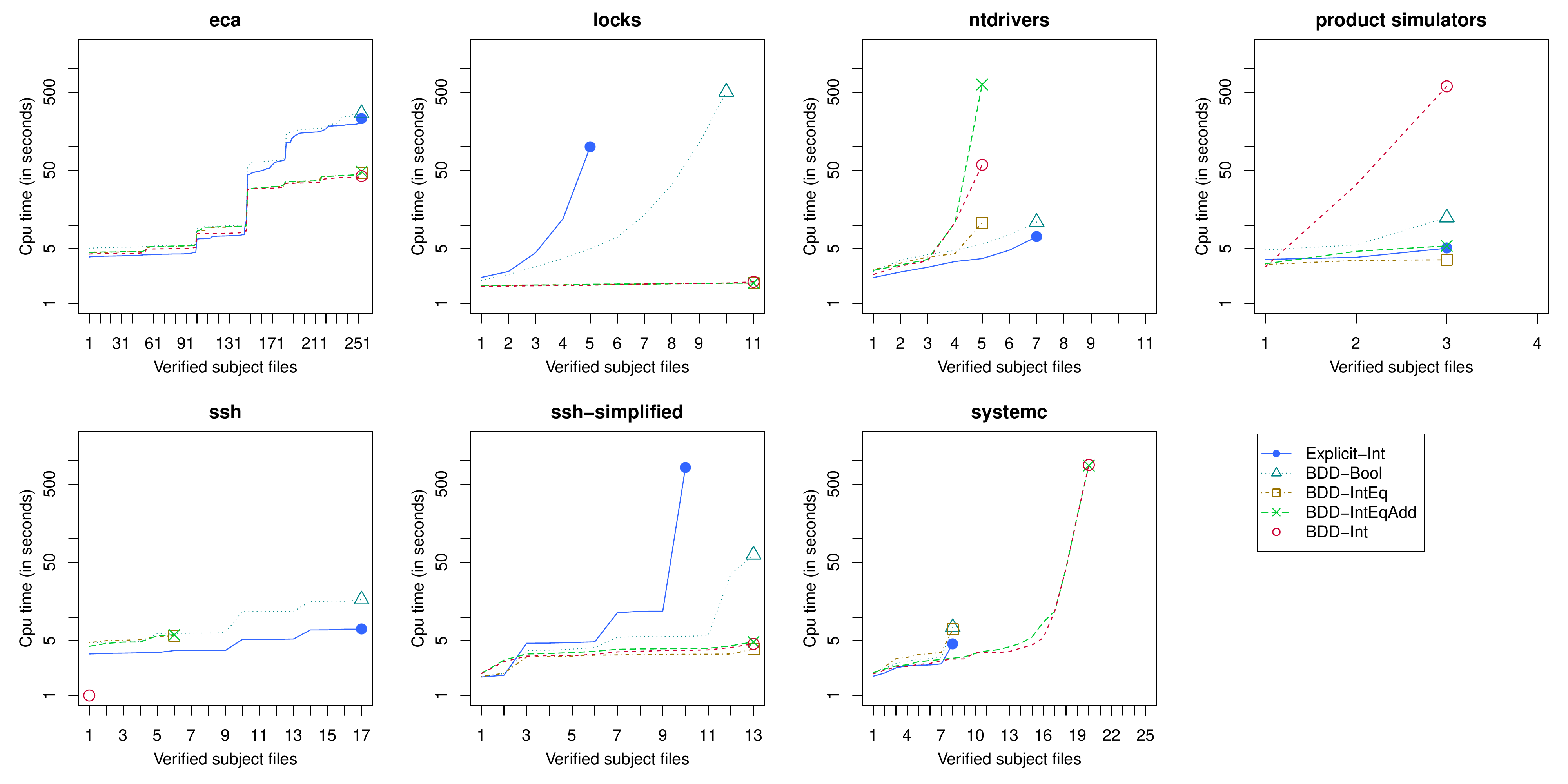}
\caption{The quantile plots show the performance of the different configurations 
per benchmark set.
Each plot shows the data of one benchmark set.
Each data point $(x,y)$ shows the $x$-fastest verification run which needed $y$ seconds.
All plots have logarithmic $y$-axes.
In the \textsc{ssh} benchmark set, the configuration \textit{BDD-Int} could not solve
any verification task.
}
\label{fig:QPlotsPerBenchmark}
\end{figure*}

\subsection{Verification Tasks}
\label{sec:subject_systems}
We evaluate our approach on 7 benchmark sets that, in total, consist of 335 verification tasks 
to be solved.
The benchmark sets are (with number of verification tasks):\\

\begin{tabular}{ll}
\textsc{eca} (254) & \textsc{locks} (11)\\
\textsc{ntdrivers} (11) & \textsc{product simulators} (4)\\
\textsc{ssh} (17) & \textsc{ssh-simplified} (13)\\
\textsc{systemc} (25) &\\
\end{tabular}\\

All verification tasks of the benchmark sets have been used in international competitions on
 verification tools~\cite{SVCOMP12,RERS12}; they are publicly available via the 
 \cc repository\footnote{\url{http://cpachecker.sosy-lab.org/}}.
Overall, this benchmark suite for software verification 
is the most comprehensive and diverse suite of this 
kind that exists.
It covers various application domains, such as device drivers, software product lines,
and embedded-systems simulation.

For our experiments, we consider only the verification tasks with expected result `safe'
(335 verification tasks).
This choice makes it necessary that the model checker explores the whole state space.
If we also included `unsafe' verification tasks, the runtime would largely depend on
which program path is explored first.
Since we focus on differences between combinations of abstract domains, 
the influence of the path precedence algorithm would merely blur the results.

The description of the systems that follows is partly taken from the report on the 2012 competition on software verification~\cite{SVCOMP12}.
Unless stated otherwise the systems are taken from this competition.
The set \textsc{ntdriver-simplified} contains verification tasks that are based on 
device drivers from the Windows NT kernel.
The sets \textsc{ssh} and \textsc{ssh-simplified} contain verification tasks that represent 
the connection-handshake protocol between SSH server and clients.
The verification tasks \textsf{ssh-simplified} 
have been manually pre-processed to remove heap accesses.
Each file checks a protocol-specific safety property.
The set \textsf{lock} contains files from the \cc repository
that were designed to investigate scalability properties of model-checking optimizations.
The files in the set \textsc{systemc} are provided by the SyCMC project~\cite{Cimatti10} 
and were taken (with some changes) from the SystemC distribution.
The benchmark set \textsc{eca} contains event-condition-action (ECA) programs, 
a kind of systems that is often used in industry.
The files in our benchmark set have been used in the 
RERS Grey-Box Challenge 2012~\cite{RERS12} on verification of ECA systems.
The benchmark set \textsc{product simulators} has been used in the competition 
on software verification 2013.
They model the variability of some product lines\,\cite{FAV_ICSE13}.

\paragraph{Domain Types}
Considering Fig.~\ref{fig:usageTypes}, an immediate observation is 
that the number of \Int variables is usually the lowest.
An exception to this observation are the benchmark sets \textsc{ssh} and \textsc{ntdrivers},
in which the number of \Int variables is extremely high compared to all other domain types.
In the other benchmark sets, we note that there is a significant number of variables 
that do not fall into the domain type \Int.
This confirms the basic premise of our approach, that there are enough variables 
in each class of variables, in order to explore the optimization potential.
In most benchmark sets, the domain type with the largest number of variables
is either \Bool or \IntEq.
We expect that optimizations for the domain types \Bool and \IntEq pay off, 
especially in the benchmark sets \textsc{eca}, \textsc{locks}, and \textsc{systemc}, 
because these domain types cover a large part of the variables in those sets.
The benchmark set \textsc{ssh-simplified} also has a high number of \IntEqAdd variables, 
so we expect a difference between configurations that use different abstract domains for this domain type.

\subsection{Results}
\label{sec:results}
Due to the huge amount of verification results, we cannot provide the raw data of 
all verification runs in the paper.
Instead, we show results aggregated by subject systems and configurations.
We also provide a website\,\footnote{\url{http://www.sosy-lab.org/projects/domaintypes/}}, 
where all results are available in form of interactive plots.
The website does provide the raw data and the logfiles of all verification runs.

\begin{table*}[t]
\begin{center}
\fontsize{6.5pt}{7.5pt} \selectfont
\addtolength{\tabcolsep}{-0.2ex}
\begin{tabular}{lD{.}{.}{2.3}D{.}{.}{2.3}D{.}{.}{2.3}D{.}{.}{2.3}D{.}{.}{2.3}D{.}{.}{2.3}}
\hline
\phantom{\large A}
\input{experiments/CPAchecker/correctnessTable.tex}
\end{tabular}
\end{center}
\caption{Verification statistics for each configurations;
each entry states the percentage of correctly solved verification tasks 
for the given configuration.}
\label{tab:allBenchmarks}
\end{table*}

\paragraph{Effectiveness}
Table~\ref{tab:allBenchmarks} gives an overview of the number of correctly solved 
verification tasks.
Each row shows the results for one benchmark set.
For each configuration, we show which percentage of the verification tasks could be 
correctly solved.

The table suggests that some verification tasks are difficult to verify.
In the benchmark set \textsc{systemc}, most configurations solve 
less than half of the verification tasks.
Most of these failures are caused by timeouts and out-of-memory terminations;
some are also due to limitations of the implemented abstract domains.
We note that the combined configurations demonstrate very good effectiveness results.
In terms of effectiveness, there is no clear winner, which suggests 
to further investigate verification based on domain-types.

\paragraph{Efficiency}
Before we discuss the details, we briefly give an overview over the results, 
based on Fig.\,\ref{fig:QPlotsPerBenchmark}.
The diagrams show the performance of the configurations 
in separate quantile plots for each benchmark set.
A point $(x,y)$ in a quantile plot states that the $x$th-fastest verification run 
of the respective configuration took $y$\,s of CPU time.
The right-most $x$-value of a configuration indicates the total number of correctly
solved verification tasks.
The area below the graph is proportional to the accumulated verification time.

For the benchmark set \textsc{eca}, 
the configurations that encode \IntEq variables in BDDs are efficient.
The configuration \textit{BDD-Bool} performs
similar to \textit{Explicit-Int}.

The benchmark set \textsc{lock} is a very good example for extremely good performance
of BDDs encodings:
the configurations \textit{Explicit-Int} and \textit{BDD-Bool} show a significant 
growth with increasing problem size, while the other configurations 
(encoding all variables as BDDs) perform extremely well.

For the benchmark set \textsc{product simulator}, the configuration \textit{BDD-IntEq} 
is fastest, followed by the configuration \textit{Explicit-Int} and then \textit{BDD-IntEqAdd}.

The benchmark \textsc{systemc} shows a very interesting detail: 
there are only two configurations 
that are able to solve most verification tasks: \textit{BDD-IntEqAdd} and \textit{BDD-Int}.
All other configurations fail on exactly the same verification tasks.

The plots also show that our approach does not perform well on two benchmark sets, 
namely the benchmark sets \textsc{ntdrivers} and \textsc{ssh}.
On these benchmark sets, all combined configurations perform worse than the explicit analysis.
\subsection{Relating Results to Domain Types}
\label{sec:analysis}
To explain why different configurations have different performance results on different 
benchmark sets, we relate the results in 
Fig.~\ref{fig:QPlotsPerBenchmark} to the domain-type statistics in Fig.~\ref{fig:usageTypes}.

The verification tasks in benchmark set \textsc{eca} contain many variables of domain type 
\IntEq (up to 58 variables), and therefore, consistently, the configurations that 
represent \IntEq variables with the BDD domain are performing best.
This indicates that tracking \IntEq variables with BDDs is a good idea.
The performance result is in line with the results of a recent paper on 
BDD-based software model checking~\cite{CPABDD}, in which similar experiments are discussed.
Our analysis of the variables in Fig.~\ref{fig:usageTypes} explains the good result in that paper.

The benchmark set \textsc{locks} shows similar results:
The configuration \Bool shows that encoding the numerous \Bool variables in BDDs improves the
performance.
Also, encoding \IntEq variables in BDDs improves the performance further.
The \textsc{locks} benchmark set has been designed to show the negative impact of a 
growing state space in analyses that work like the explicit-value analysis
(single-block encoding without joins)~\cite{ABE};
therefore, this result was expected.

For the benchmark set \textsc{ntdrivers}, we observe the opposite
effect: the more variables are encoded in BDDs, the worse the performance gets.
Fig.~\ref{fig:usageTypes} shows that the verification tasks
in this benchmark set contain many \Int variables.
Only a few variables have a simple domain type and can be encoded in BDDs.
If this is done, the BDD domain has a negative impact on the overall performance:
the configurations \textit{Explicit-Int} performs better than the BDD-based configurations.

The verification tasks in benchmark set \textsc{product simulator}
contains many \IntEq (130-256) and \IntEqAdd variables (56-96).  The
performance of the configuration \textit{BDD-IntEq} is best, as
expected.  However, the performance of the configuration
\textit{BDD-IntEqAdd} is worse than or equal to the performance of the
configuration \textit{Explicit-Int}.  This means that it is too
expensive to represent all variables of type $IntEqAdd$ in the BDD.
We have to use 32 BDD variables for each program variable and encode
each operation on the program variables as BDD operations (e.g.,
additions with a full adder).  This involves more complex operations
on the BDDs, compared to the \IntEq variables, for which the
operations are simple assignments and comparisons.  The benchmark also
shows that the configuration \textit{BDD-Int} needs substantially more
time than all other configurations on two of the verification tasks,
which means that there are some \Int variables which cannot be handled
properly in the BDD domain.  This configuration can still solve the
verification tasks, however, it has to spend more time on expensive
BDD operations and is therefore slower.

Similar to the benchmark set~\textsc{ntdrivers}, the configuration \textit{Explicit-Int}
is the best configuration for the benchmark set \textsc{ssh}. 
These two benchmark sets do not benefit from using the BDD domain.
The configurations that handle more variables than the variables contained in \Bool with BDDs,
cannot solve many of the verification tasks in the benchmark set.
The verification tasks contain many variables of domain type~\Int.

The verification tasks of the benchmark set \textsc{ssh-simplified} contains
some variables of each domain type, but not too many overall.
Most of the variables are of domain types \Bool and \IntEqAdd, and this is reflected by the 
good performance of the configurations \textit{BDD-IntEq} and \textit{BDD-IntEqAdd}.
The benchmark set \textsc{ssh-simplified} has been derived from the files in the 
benchmark \textsc{ssh} by removing heap access operations.
The effect on the variable setup is visible in Fig.~\ref{fig:usageTypes}, 
where in \textsc{ssh} nearly all variables are of type \Int, and in \textsc{ssh-simplified} 
most variables are of simpler domain types.
This has a large impact on the effectiveness and performance of the verification:
The more variables are encoded in BDDs, the better is the performance.
A closer look at the source code of the verification tasks
reveals that during the ``simplification'', some local variables 
were made global.
The explicit domain is inefficient in tracking all these variables, as can be seen in 
Fig.~\ref{fig:QPlotsPerBenchmark}.
If these variables are encoded in BDDs, these problems do not exist.

The benchmark set \textsc{systemc} has a quite interesting characteristic: 
it contains many \IntEq variables, but the most successful configurations are 
\textit{BDD-IntEqAdd} and \textit{BDD-Int}.
The reason for this is that many of the verification tasks
contain uninitialized variables with an inequality 
comparison and an addition on one of the variables.
The explicit-value analysis is not good at tracking facts in such cases,
and thus, the configurations that use the explicit-value analysis 
for some variables terminate unsuccessfully.
The configurations \textit{BDD-IntEqAdd} and \textit{BDD-Int} 
can efficiently analyze all operations and solve the verification tasks successfully.

\subsection{Evaluation}
Our experimental study has shown that the performance of combined approaches
depends inherently on the domain types of the variables in the program.
If the verification tasks contain variables of domain type \textit{IntEq}, 
then representing these 
variables with the BDD domain can improve the performance significantly.
If the programs contain some \IntEqAdd variables, analyzing them with the BDD domain 
does also improve performance (e.g., in the benchmark \textsc{ssh-simplified}).
However, there seems to be a threshold at about 200 variables, above which the 
overhead of analyzing integer operations with BDDs has a visible negative effect.

Analyzing variables of domain type \textit{IntEq} using BDDs 
is much more efficient, so we expect 
the threshold to be considerably higher for this domain type.
In general, it is not (yet) possible to make a clear statement about these thresholds,
because the performance also depends on the other variables 
and on the operations that are used in the program.
To counter the negative effect of large sets of variables of ``expensive'' domain types, 
we will in the future investigate three techniques:
(1)~introduce further detailed domain types such that
we can make more fine-grained decisions on the domain assignment,
(2)~a prioritization function to assign only the ``most profitable'' 
variables to the respective domain, and
(3)~introduce several instances of the abstract domains in the program analysis, where each 
instance handles an independent partition of the domain type.

\begin{figure}[t]
\centering
\vspace{-8mm}
\includegraphics[width=\columnwidth]{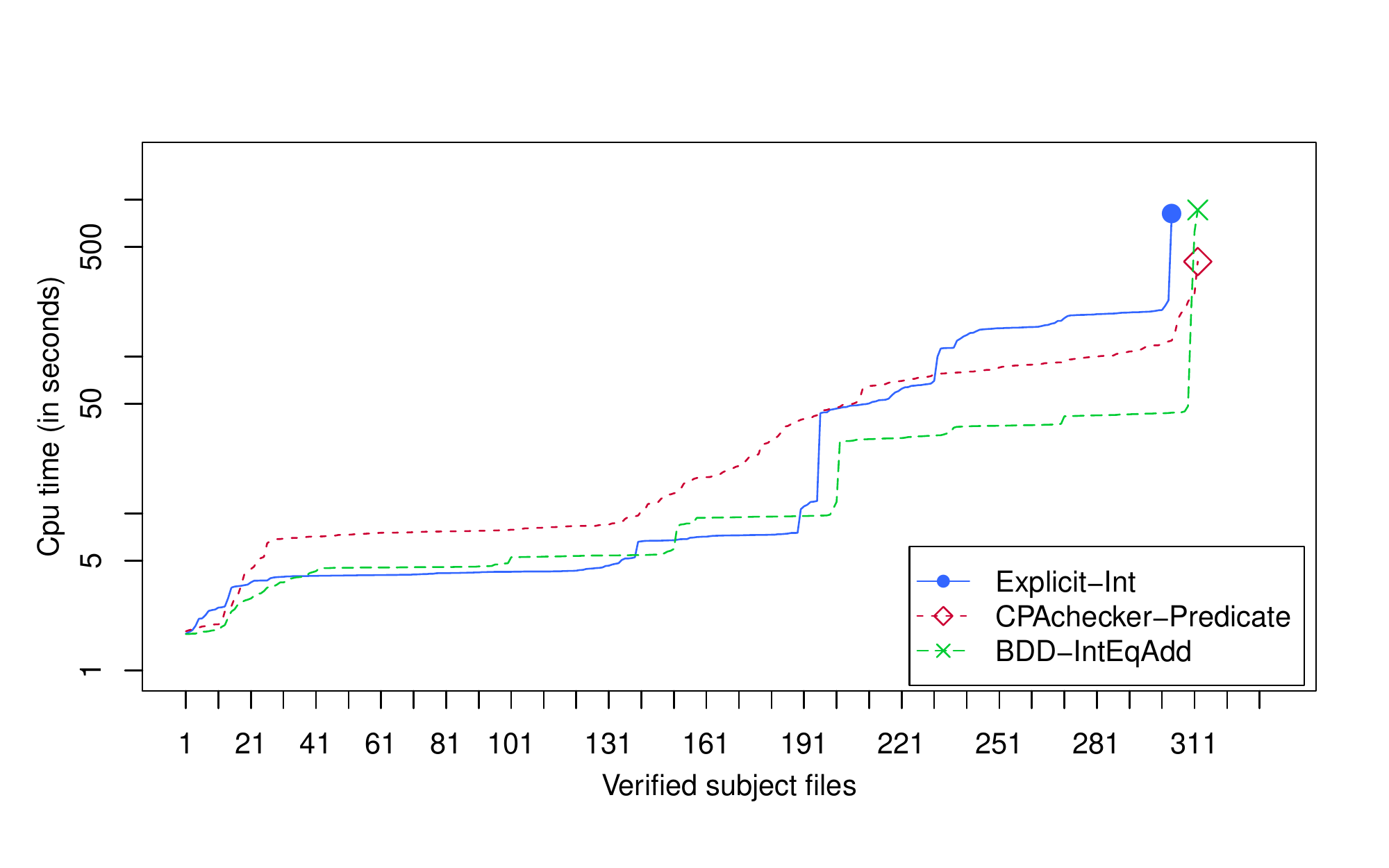}
\caption{Performance of the configuration \textit{BDD-IntEqAdd}
with the baseline configuration~\textit{Explicit-Int} and a competitive 
implementation of predicate analysis; quantile plots}
\label{fig:QPlotIntAddvsPred}
\end{figure}

To give further evidence for the quality of our approach, we compared the 
configuration \textit{BDD-IntEqAdd} with \cc-\textsc{Predicate},
which is based on a completely different abstract domain (predicate abstraction)
and CEGAR, interpolation, and adjustable-block encoding~\cite{ABE}.
The configuration \cc-\textsc{Predicate} has won the category `ControlFlowInteger' 
in the SV-COMP~2012 and another analysis, based on \cc-\textsc{Predicate}, 
has won the category `Overall' in the SV-COMP 2013.
Therefore, this experiment is a representative comparison with the state-of-the-art
in performance.
Fig.~\ref{fig:QPlotIntAddvsPred} shows a quantile plot comparing the 
performance of the new configuration~\textit{BDD-IntEqAdd} against \cc-\textsc{Predicate}.
Both approaches can successfully solve most verification tasks, however, 
the new approach performs better in terms of CPU time, 
compared to both \cc-\textsc{Predicate} and configuration~\textit{Explicit-Int}.
In terms of solved verification tasks, \cc-\textsc{Predicate} performs
slightly better.

\paragraph{Evaluative Summary}
To evaluate our approach, we briefly discuss the statements that we list at the 
beginning of the section, based on the results.
The first statement, concerning the domain types has already been discussed 
(Fig.~\ref{fig:usageTypes}).
Concerning the variable partitioning, we confirm that analyzing variables 
of different domain types with different abstract domains can make a huge difference,
in terms of effectiveness and efficiency.
There are several new configurations that outperform the existing configurations
(only explicit-value domain or only BDD domain) on several benchmark sets.
The benchmark sets \textsc{ntdrivers} and \textsc{ssh} mainly consist of 
variables that do not fit in simple domain types.
So the best domain for these files is in fact the previously existing 
configuration~\textit{Explicit-Int}.
On the other benchmarks, we can assign the variables according to their 
domain types to abstract domains and in nearly all cases,
the configuration that encodes the dominant variables with BDDs and uses the 
explicit-value domain for the rest performs best.
The configuration \textit{BDD-Int} performs nearly as good as the combined 
configurations, however, on the verification tasks from \textsc{product simulators} and \textsc{ssh} 
it is apparent that including the support of the explicit analysis for 
\Int variables is critical.
Our comparison with the configuration \cc\textsc{-Predicate} shows that our combined 
approach is also competitive against other, more optimized verification approaches
that use the full power of existing technology (including CEGAR, which we eliminated 
from the discussion to not blur the picture).
Overall, it might be beneficial to use 
the BDD domain for variables of domain type \emph{IntEqAdd}, rather than using the 
explicit domain.
This is confirmed by the performance of configurations \textit{BDD-IntEq}, 
\textit{BDD-IntEqAdd}, and \textit{Explicit-Int} in most benchmark sets 
(except for (\textsc{ntdrivers} and \textsc{ssh}).
The configuration~\textit{BDD-IntEqAdd} can successfully solve many verification tasks 
of the set \textsc{systemc}, for which all other configurations fail.

\subsection{Threats to Validity}
\label{sec:threats}
\paragraph{Threats to Internal Validity}
Due to the large number of benchmark verification tasks, 
we could not execute every verification run several times in order to perform statistical 
significance tests on the results.
However, we argue that relevant parts of our discussion do not depend on 
the timing results (e.g., successfully solved verification tasks; 
number of variables per domain type)
and that the performance results are convincing.
We performed several minor changes to the benchmark sets and configurations 
to address technical difficulties until we had the final setup.
None of these changed the big picture of the results.

\paragraph{Threats to External Validity}
The major threat to external validity is that we have used only two 
abstract domains (explicit-value and BDD) to distribute variables.
However, these domains are perfectly suited for the chosen domain types and 
complement each other very well.
Therefore it was an intuitive choice.
Combinations with other domains have to be explored in future work.
Another threat is that our benchmark set consists of a small number of verification tasks or of 
too many verification tasks of a certain kind.
We argue that the benchmark sets have been used by well-respected international 
competitions and represent programs that are used for evaluation by others researchers.
Also, we have seen that these programs contain a diverse set of variables.

%% file: experiments/CPAchecker/correctnessTable.tex
 & \multicolumn{1}{c}{\textit{Explicit-Int}} & \multicolumn{1}{c}{\textit{BDD-Bool}} & \multicolumn{1}{c}{\textit{BDD-IntEq}} & \multicolumn{1}{c}{\textit{BDD-IntEqAdd}} & \multicolumn{1}{c}{\textit{BDD-Int}}\\
\hline
\textsc{eca}                & 100 & 100 & 100 & 100 & 100\\
\textsc{locks}              & 45 & 91 & 100 & 100 & 100\\
\textsc{ntdrivers}          & 64 & 64 & 45 & 45 & 45\\
\textsc{product simulators} & 75 & 75 & 75 & 75 & 75\\
\textsc{ssh}                & 100 & 100 & 35 & 35 & 0\\
\textsc{ssh-simplified}     & 77 & 100 & 100 & 100 & 100\\
\textsc{systemc}            & 32 & 32 & 32 & 80 & 80\\
\hline

%% file: sections/related.tex
\section{Related Work}
The two symbolic domains BDDs and Presburger formulas were previously use
as representation for boolean and integer variables~\cite{Bultan00}.  
The approach was evaluated on two systems, a
control software for a nuclear reactor's cooling system and a
simplified transport-protocol specification.  
In contrast to our work,
the approach is not based on a separate analysis to determine domain types of variables, but
include the type analysis in the actual model-checking process.  
By performing the domain-type analysis in advance, we avoid overhead
during the model checking process.  
The approach was evaluated on a much smaller benchmark set.

We infer domain types for program variables
according to their usage in program oparations.
This principle is also used by
for type- and memory-safety analysis
of C programs with \emph{liquid types}~\cite{Rondon2012}.  
A static program analysis is used to determine for each variable a predicate
that restricts the possible values of the variable (the \emph{liquid type}).  
In a second step, each usage of the variable is checked for
type-safety, or if it could lead to an unsafe memory access.  
In contrast to domain types, \emph{liquid types} use a predicate for each variable. 
\emph{liquid types} are fine-grained.  
Domain types can be seen as coarse-grained in comparison,
but the granularity is flexible in both approaches.
Our type checker for domain types does not depend on an SMT solver.

\emph{Roles of variables} are used to analyze 
programs submitted by students~\cite{Bishop05}.
Program slicing and data-flow analysis is applied to determine the role of 
each variable (e.g., \emph{constant} or \emph{loop index}).
The role determined by the analysis is compared to the role 
that the students have assigned to the variables.
This work falls into the area of automated program comprehension.
The rather strong behavioral variable types might be 
interesting to extend our work.

Java Pathfinder~\cite{VHB+03} has an extension
that combines the standard explicit 
analysis with a BDD-based analysis for boolean variables~\cite{RAR11,FAV_ICSE13}.
In that approach, we manually selected the variables that are to be tracked by BDDs, 
based on domain knowledge.
Our new approach handles a broader set of domain 
types and categorizes them automatically.

\textsc{Bebop}~\cite{Ball00}, a model checker for boolean programs,
encodes all program variables (only booleans in this case) in BDDs, 
and use explicit-state exploration to handle the program counter.
Our domain-type analysis would correctly classify all variables as \Bool
and encode them with BDDs; thus, we subsume this approach as a special instance.
A similar strategy was followed by others~\cite{CimattiRB01}.

A hybrid approach combining explicit and BDD-based representations
analyzes the program variables with BDDs 
and the states of the property automaton explicitly~\cite{Sebastiani05}.
In our setting, this translates to encoding all program variables
in BDDs, because the property automaton runs separately in parallel, in \cc.
This case can be represented in our general framework
as configuration \emph{BDD-Int}.

%% file: sections/conclusion.tex
\section{Conclusion}
We introduced the concept of \emph{domain types}, which makes it possible
to assign variables to different abstract domains based on their usage by program operations. 
In this paper, we outlined the approach by introducing a type hierarchy that splits
the declared type `integer' into four more detailed domain types,
which reflect the usage of variables in the program.
We performed an experimental study with two abstract domains, in order to confirm that the
domain assignment based on domain types has a significant impact on the effectiveness
and efficiency of the verification process.
In the experiments, we considered five domain assignments:
one for each considered abstract domain that tracks all program variables
in one single abstract domain, without considering the different domain types,
and three with different assignments of the variables to the two abstract domains
according to the domain type.

The insight of our work is that the concept of domain types is a simple yet powerful
technique to create verification tools that implement a better choice for the domain assignment
--- state-of-the-art is to use either one single abstract domain, or a fixed combination
of abstract domains that adjust precisions via CEGAR or otherwise dynamically, during the 
verification run.
We confirmed that the benchmark set contains a significant set of variables for which we can determine 
different, narrower domain types.
The insight of the domain type \emph{IntEq} is that this domain type 
(and even more its sub-type \emph{Bool})
dramatically decreases the number of possible values of the variables in the internal representation,
and thus can yield a large speedup in verification time.
The experiments show that performance can be improved if the variables are 
tracked in an abstract domain that is suitable for the domain type of the variable.

There is still room for improvement.
For example, we can consider CEGAR as an orthogonal improvement of the overall verification
configuration.
We can also combine more analyses through the communication between 
different domains using the strengthen operator --- so far we use a simple cartesian combination.
This would enable even finer domain assignments and to use more abstract domains in parallel.